\documentclass[%
superscriptaddress,
 amsmath,amssymb,
 aps,
prc,
twocolumn,
]{revtex4-2}
\usepackage{graphicx}
\usepackage{dcolumn}
\usepackage{bm}
\usepackage{hyperref}
\usepackage[table]{xcolor}
\usepackage{array}
\usepackage{lineno}
\usepackage{blindtext}
\usepackage{enumitem}
\usepackage{booktabs}

\hypersetup{breaklinks=true, colorlinks=true, citecolor=cyan, linkcolor=blue, urlcolor=blue,filecolor=blue}

\begin{document}
\preprint{APS/PRC}


\title{Hofstadter-Herman Visualization as a Diagnostic Tool for Systematic Effects in Electromagnetic Form Factor Extractions}

\author{Tyler Williams}
\affiliation{Department of Physics, Duquesne University, Pittsburgh, PA 15282}
\author{Jennifer Rittenhouse West}
\affiliation{INFN Turin Section, Turin, Italy 10125} 
\affiliation{Department of Physics, University of Turin, Turin, Italy 10125}
\author{D. W. Higinbotham}
\affiliation{Thomas Jefferson National Accelerator Facility, Newport News, VA 23606}
\author{Fatiha Benmokhtar}
\email{Contact person: benmokhtarf@duq.edu}
\affiliation{Department of Physics, Duquesne University, Pittsburgh, PA 15282}

\date{\today}

\begin{abstract}
The internal charge and magnetization distributions of the proton are characterized by electromagnetic form factors $G_E$ and $G_M$. They are extracted from elastic electron-proton cross sections measured at fixed momentum transfer over a range of beam energies and scattering angles via Rosenbluth separation. Conventionally, form factor values are obtained by plotting reduced cross sections against the virtual photon polarization parameter $\epsilon$ and then extracting the slope and intercept of the best-fit lines. An alternative visualization method, proposed by Hofstadter and Herman in 1960, plots form factors directly instead. The values of $G_E^2$ and $G_M^2$ are immediately visible from the intersection region of the curves and their uncertainty bands. In this work, we apply Hofstadter-Herman visualization to classic 1994 SLAC elastic scattering data. The 1994 data include a $\sim$4\% downward normalization adjustment applied to the 1.6 GeV/$c$ spectrometer measurements. Overlaying modern global fit extractions on Hofstadter-Herman plots, we show that intersection regions obtained without including this adjustment sit closer to the global fit values. Such tensions are not readily visible in conventional plots. Our results motivate adopting this visualization method as a routine diagnostic cross-check at the Electron–Ion Collider and elsewhere, to flag normalization shifts and related adjustments before they enter global fits.
\end{abstract}

\maketitle

\section{Introduction}
Nucleon electromagnetic form factors characterize how electric charge and magnetization are distributed within the proton and neutron and are among the most direct experimental probes of their internal structure \cite{Ye:2017gyb,Lin:2021xrc,Alberico:2008sz,A1:2013fsc,Kubon:2001rj}. Experimentally, form factors are extracted from elastic scattering of charged leptons off nucleon targets. Liquid hydrogen targets are typically used for proton measurements, while neutron information is obtained from quasielastic scattering on light nuclei such as deuterium or $^3$He. Theoretically, Quantum Electrodynamics (QED) shows that elastic electron–nucleon scattering proceeds, at leading order, through the exchange of a single virtual photon carrying four–momentum $q$ with invariant four–momentum transfer squared defined as $Q^2\equiv -q^2$ \cite{Rose:1948zz,Rosenbluth:1950yq,Sachs:1962zzc}.  For more than 70 years, elastic $ep$ scattering has played a central role in the study of the internal structure of the proton \cite{Hofstadter:1955ae,Perdrisat:2006hj,Hofstadter:1960}.

Elastic scattering from a spin-$\frac{1}{2}$ nucleon requires two independent Lorentz–invariant form factors to describe its electromagnetic current. These are expressed as the Sachs electric and magnetic form factors, $G_E(Q^2)$ and $G_M(Q^2)$. In 1962 Sachs showed that, in the Breit frame and in the non-relativistic limit, $G_E(Q^2)$ and $G_M(Q^2)$ can be interpreted as Fourier transforms of the nucleon’s spatial charge and magnetization densities \cite{Sachs:1962zzc}. The unpolarized elastic scattering cross section depends on both form factors and their relative weights vary with scattering angle and beam energy at fixed $Q^2$.

The standard procedure for separating $G_E$ and $G_M$ from unpolarized cross sections is the Rosenbluth (or longitudinal–transverse) separation. Measurements are taken at fixed $Q^2$ but with different combinations of beam energies and scattering angles. The data are then typically plotted as reduced cross sections versus the virtual photon polarization parameter $\epsilon$ (a known kinematic function of $Q^2$ and $\theta$, see Section \ref{sec:theory}). A linear fit through the data allows for an extraction of $G_E^2$ and $G_M^2$ from the slope and intercept of the line. Although robust in principle, this method can obscure systematic inconsistencies between data sets due to the fitting of data to a line and subsequently extracting the form factors. 

An alternative visualization was proposed by Hofstadter and Herman in 1960 \cite{Hofstadter:1960}. Instead of plotting reduced cross sections, they plotted data directly in form factor space, where each data point with uncertainties at fixed $Q^2$ corresponds to an allowed band of form factor values. When multiple measurements at the same $Q^2$ but with different combinations of beam energy and scattering angle are displayed together, the overlapping region of these bands identifies the consistent form factor pair.  Hofstadter and Herman originally presented their visualization in the Dirac-Pauli form factor plane $(F_1,F_2)$. We apply the same concept in the Sachs basis $(G_E^2,G_M^2)$ \cite{Sachs:1962zzc}. 

In this work, we apply both the conventional Rosenbluth plotting method and the Hofstadter–Herman visualization to the 1994 SLAC elastic–scattering data of Andivahis \textit{et al.} \cite{Andivahis:1994}. We show that the Hofstadter–Herman approach reveals tensions in the data, most notably between normalization-adjusted and unadjusted data sets, that are not apparent in traditional Rosenbluth plots. Comparison with Jefferson Lab Global Fits \cite{Puckett:2017} suggests that the SLAC-applied $\sim$4\% downward normalization correction on a subset of the data does not improve agreement with the 2017 global fit extractions. This study highlights the diagnostic power of visualizing data directly in form factor space to uncover systematic effects and data set inconsistencies, which is relevant for the Electron-Ion Collider \cite{Accardi:2012qut,Achenbach:2023pba}.

The remainder of this paper is organized as follows. Section II reviews elastic electron–proton scattering and electromagnetic form factors and develops the conventional Rosenbluth and Hofstadter–Herman representations used in our comparison. Section III reanalyzes the 1994 SLAC data of Andivahis et al. \cite{Andivahis:1994}, using both representations to examine the effects of the normalization adjustment applied to the 1.6~GeV/$c$ spectrometer data. Section IV discusses future applications of the Hofstadter–Herman method and summarizes our conclusions. Appendix A details the treatment of experimental uncertainties.

\section{\label{sec:theory}Theoretical Foundations}
Electromagnetic form factors provide a quantitative description of the spatial distributions of electric charge and magnetization within the proton. The electric form factor, $G_E(Q^2)$, reflects the electric charge distribution, while the magnetic form factor, $G_M(Q^2)$, characterizes the magnetization profile. Both are functions of the squared four-momentum transfer, $Q^2$, and are fundamental to understanding the internal structure of the nucleon \cite{Ye:2017gyb,Alberico:2008sz,Lin:2021xrc}.
Form factors are directly related to the underlying quark dynamics, as the proton’s electric charge is carried by its constituent quarks and antiquarks. By analyzing the behavior of $G_E$ and $G_M$ over a range of $Q^2$ values, one can infer information about the spatial configuration and behavior of confined quarks, as well as estimate key properties such as the proton’s radius \cite{Higinbotham:2015rja,Karr:2020wgh} and shape \cite{Arrington:2011kb,Hand:1963zz}. 

As mentioned in the introduction, elastic electron-proton scattering forms the experimental basis for determining the proton’s electromagnetic form factors \cite{Perdrisat:2006hj}. In this process, an incident electron beam of known energy interacts with a stationary proton target, resulting in a scattered electron and a recoiling proton. Crucially, no energy is transferred to the internal degrees of freedom of the proton; while the proton recoils with kinetic energy, it remains in its ground state, distinguishing elastic from inelastic scattering, where the proton can be excited to resonances or break apart.

In the single photon exchange approximation, the unpolarized differential cross section can be written in the form \cite{Rosenbluth:1950yq}

\begin{align}
\frac{d\sigma}{d\Omega}
&= \left(\frac{d\sigma}{d\Omega}\right)_{\text{ns}}
\left[
\frac{G_E^2(Q^2)+\tau G_M^2(Q^2)}{1+\tau}
\right. \notag\\[-2pt]
&\left.\qquad
+\, 2\tau\, G_M^2(Q^2)\tan^2\!\left(\frac{\theta}{2}\right)
\right].
\label{eq:sigma_elas}
\end{align}

Here, $\left( d\sigma/d\Omega\right)_{\text{ns}}$ is the cross section for a point-like spinless target of electric charge $+e$  (i.e., ``no structure,'' also known as Mott scattering), $M_p = 0.938$ GeV is the proton mass, $\tau = Q^2 / 4M_p^2,$ and $\theta$ is the electron scattering angle. We work with natural units in which $\hbar = c = 1$. This formulation is based on the Rosenbluth separation technique, which allows independent extraction of electric ($G_E$) and magnetic ($G_M$) form factors by systematically varying the scattering angle while keeping $Q^2$ fixed and will be described in detail in the following subsection.

In the Rosenbluth separation experiments performed at SLAC, high-energy electron beams were directed at liquid hydrogen targets. The angular distribution of scattered electrons was measured using precision magnetic spectrometers (specifically, 1.6 GeV/$c$ and 8 GeV/$c$ spectrometers), enabling accurate determination of differential cross sections over a broad range of momentum transfers.

\subsection*{Conventional Rosenbluth Visualization}
Form factors are typically extracted using the Rosenbluth separation technique \cite{Rosenbluth:1950yq}.  This method involves determining the differential cross section at fixed $Q^2$ from data taken at different beam energies and scattering angles.  Data are conventionally plotted as the reduced cross section $\sigma_R$ versus virtual photon polarization $\epsilon$ at fixed $Q^2$. In this case, the slope corresponds to $G_E^2$ and the intercept to $\tau G_M^2$, allowing both form factors to be algebraically extracted from a straight-line fit.

The reduced cross section is defined as 
\begin{equation}
\sigma_R = \epsilon (1+\tau)\frac{d\sigma/d\Omega}{\sigma_{\text{ns}}}
\end{equation}
where the no-structure (Mott) cross section is given by
\begin{equation}
\sigma_{\text{ns}} = \frac{\alpha^2 \cos^2\left(\frac{\theta}{2}\right)}{4E^2 \sin^4\left(\frac{\theta}{2}\right)}  \frac{E'}{E},
\end{equation}
and $\epsilon$ is given by
\begin{equation}
\epsilon = \frac{1}{1 + 2(1+\tau)\tan^2\left(\frac{\theta}{2}\right)},
\end{equation}
and $E$ and $E'$ denote the incident and scattered electron energies in the laboratory frame. The reduced cross section is then divided by the square of the magnetic dipole form $G_D^2$, resulting in the following equation
\begin{equation}
\label{eq:complicated_cross}
\frac{\sigma_R}{G_D^2\left(Q^2\right)} \equiv\left(\frac{d \sigma}{d \Omega}\right) \frac{(1+\tau)}{5.18} \epsilon\frac{E^3}{E^{\prime}} \frac{\sin ^4\left(\frac{\theta}{2}\right)}{\cos ^2\left(\frac{\theta}{2}\right)}\left(1+\frac{Q^2}{0.71~{\rm GeV^2}}\right)^4
\end{equation}
where $G_D\left(Q^2\right)=\left(1+\frac{Q^2}{0.71~{\rm GeV^2}}\right)^{-2}$ is the conventional dipole parametrization used as a reference in the 1994 analysis. We note that $G_D$ is used only to scale the conventional plot and does not affect the values of the extracted form factors.

\begin{figure}[!htpb]
\centering
\includegraphics[width=\linewidth]{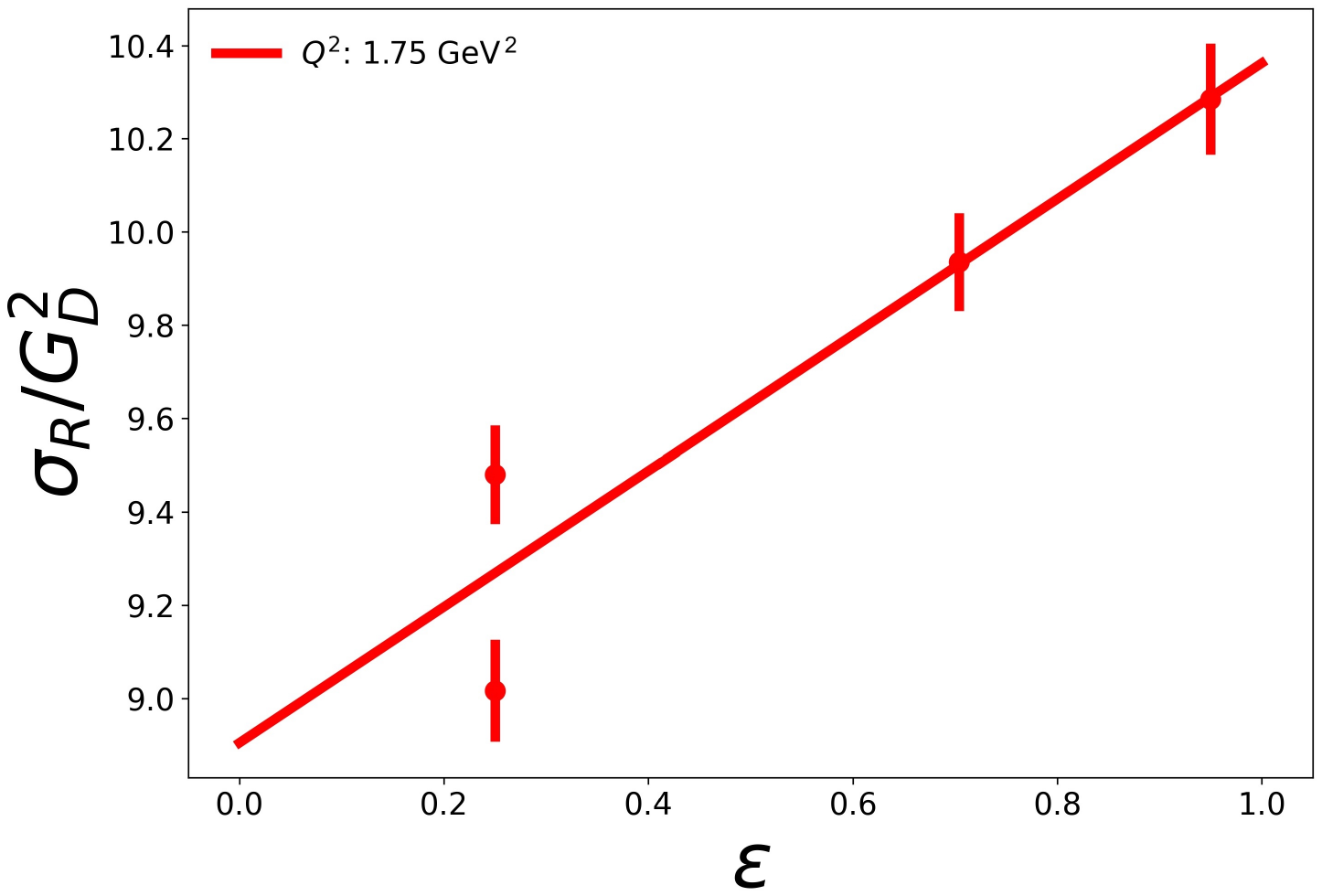}
\caption{An example of a standard Rosenbluth separation method plot at $Q^2=1.75$ GeV$^2$.}
\label{example}
\end{figure}

We use this equation in our example of conventional Rosenbluth plots to illustrate the method used in 1994 by SLAC \cite{Andivahis:1994}, shown as $\sigma_R/G_D^2$ vs. $\epsilon$ in Figure \ref{example}. The electromagnetic form factors obtained in this method are obtained from the slope and y-intercept of the best-fit line.

The different choices of plotting variables and form factor extractions emphasize that conventional visualization methods can be confusing and visually obscure the physical parameters. 
Reducing the data to a fitted slope and intercept can obscure potentially important point-to-point information.

\subsection*{Hofstadter-Herman Visualization Method}
A geometric visualization method was proposed by Richard Hofstadter and Robert Herman in 1960 \cite{Hofstadter:1960}. They suggested that the form factor data could be displayed more effectively by plotting constraint regions in form factor space, with uncertainty bands drawn around each form factor curve. The intersection region of the bands gives the form factor values consistent with all measurements. 

To derive the relationship between $G_E^2$ and $G_M^2$ for the Hofstadter-Herman plots, it is useful to rewrite Eq.~\ref{eq:sigma_elas} to make the dependence on the virtual photon polarization parameter $\epsilon$ explicit. 
Rearranging to isolate the $\epsilon$-dependence yields
\begin{equation}
\frac{d\sigma}{d\Omega} = \frac{\sigma_{\text{ns}}}{\epsilon(1+\tau)} \left( \tau G_M^2(Q^2) + \epsilon G_E^2(Q^2) \right).
\end{equation}

Substituting and rearranging algebraically, we find
\begin{equation}
\sigma_R =  \tau G_M^2(Q^2) + \epsilon G_E^2(Q^2) 
\end{equation}
which yields
\begin{equation}
G_M^2(Q^2) = \frac{\sigma_R}{\tau} - \frac{\epsilon}{\tau} G_E^2(Q^2). 
\label{eq:gm}
\end{equation}
Thus, for each measurement at fixed $Q^2$, the central measurement line has slope $-\epsilon/\tau$ and $G_M^2$-axis intercept $\sigma_R/\tau$.

\begin{figure}[!htpb]
\centering
\includegraphics[width=1.0\linewidth]{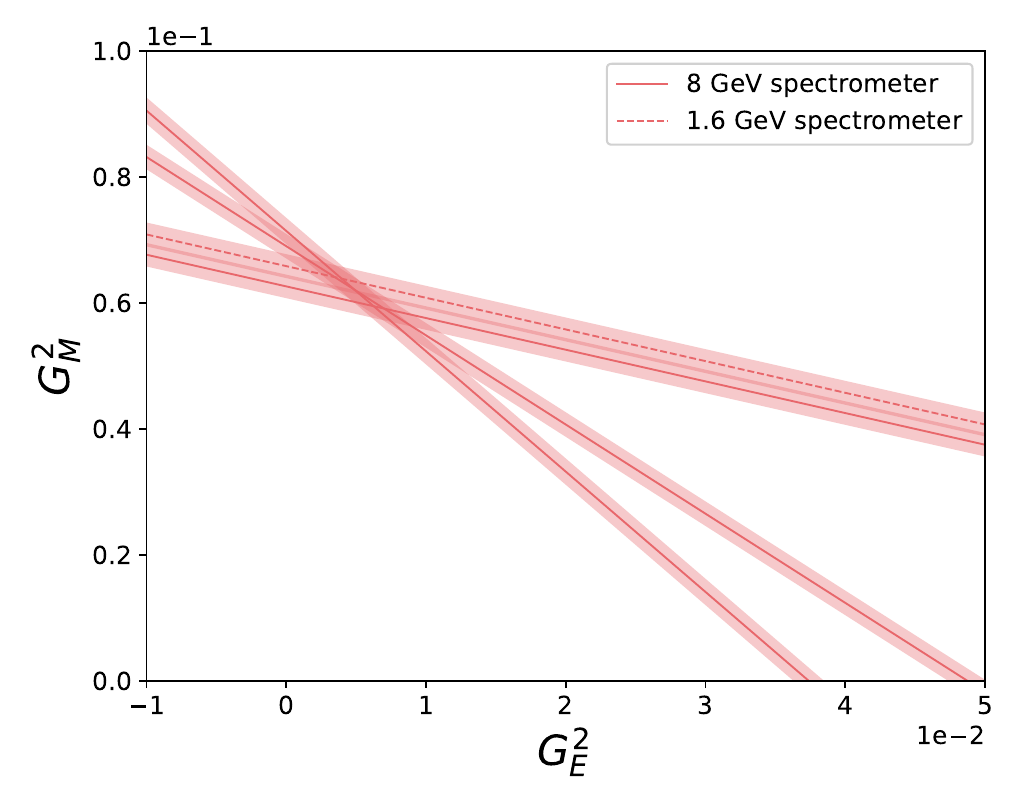}
\caption{An example of the Hofstadter-Herman visualization method at $Q^2=1.75$ GeV$^2$.}
\label{hofstadterexample}
\end{figure}

In the original Hofstadter-Herman construction, each measurement defined an ellipse in the Dirac-Pauli form factor plane $(F_1, F_2)$; the intersection of several such ellipses identified the consistent form factor pair. In the Sachs representation used here, those loci become straight bands in $\left(G_E^2, G_M^2\right)$ space, as can be seen from Eq.~\ref{eq:gm}, where each measurement at a given $Q^2$ is plotted as a line with its experimental uncertainty band. The overlap region of the bands identifies the set of ($G_E^2$, $G_M^2$) pairs that are compatible with all measurements at the same $Q^2$ (and also the magnitudes $|G_E|$ and $|G_M|$ for unpolarized scattering). Figure~\ref{hofstadterexample} shows an example Hofstadter-Herman visualization plot. 

\begin{figure*}[t]
\centering
\includegraphics[
  width=0.49\textwidth,
  trim=30 10 30 10,
  clip]{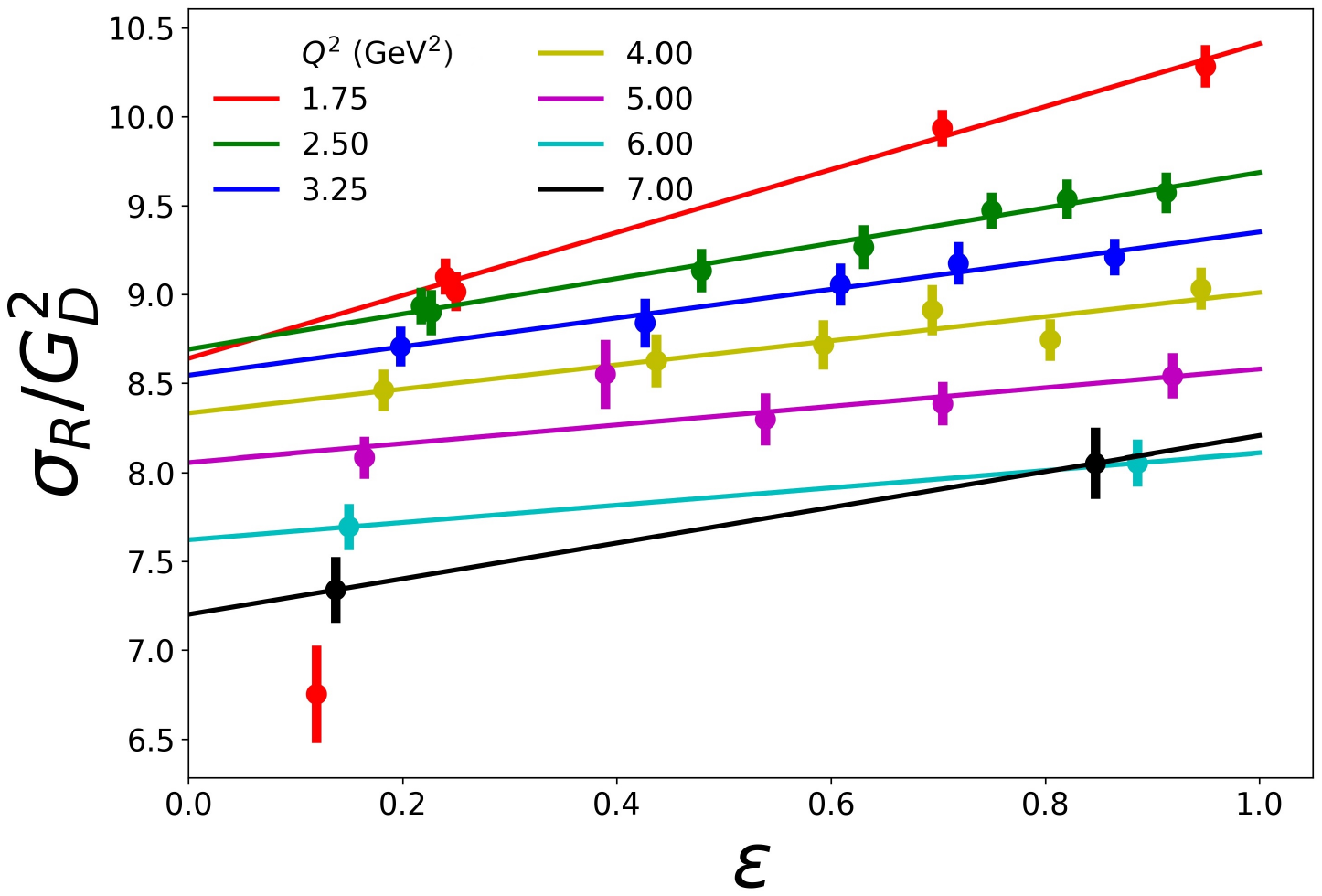}\hfill
\includegraphics[
  width=0.49\textwidth,
  trim=30 10 30 10,
  clip]{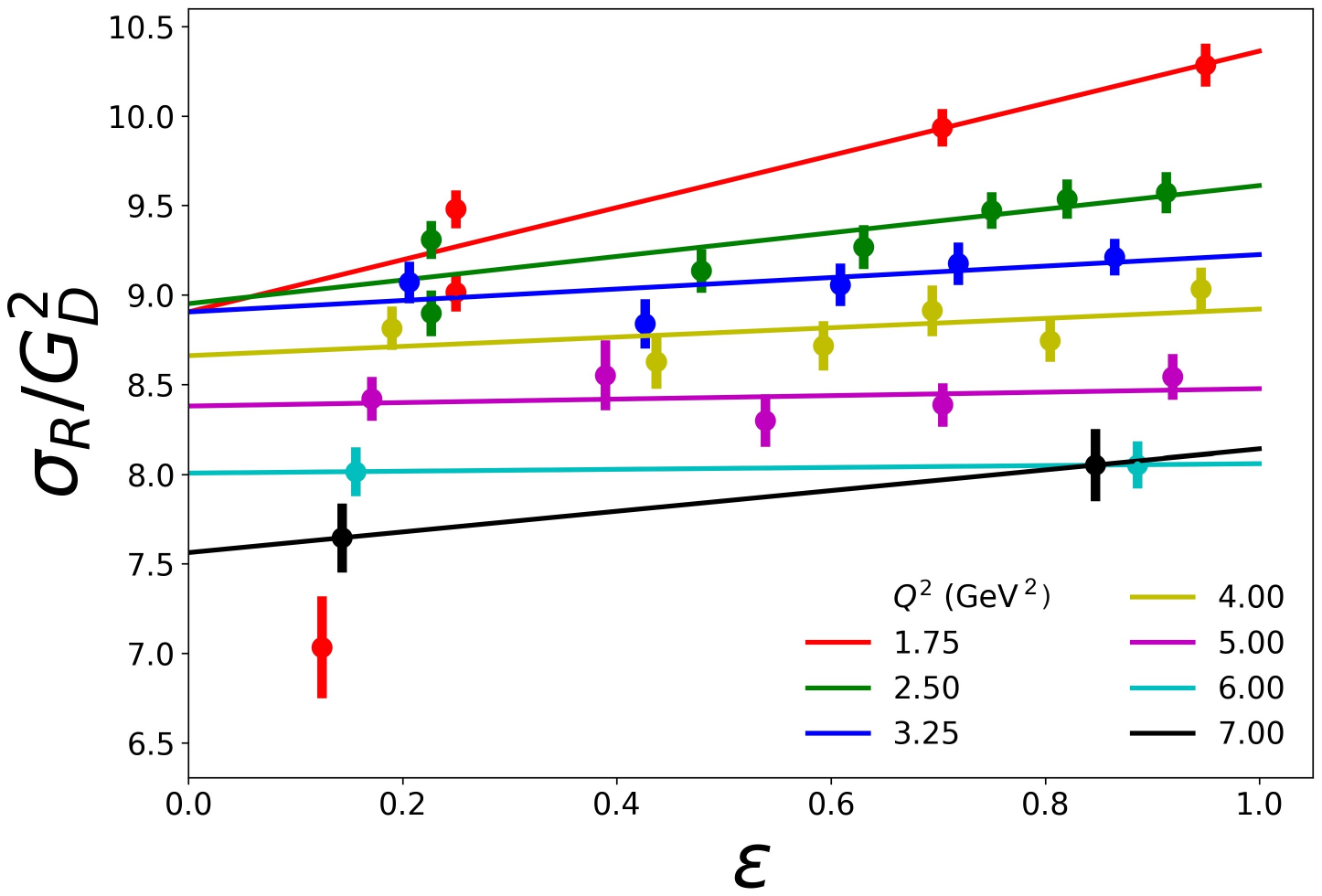}
\makebox[0.49\textwidth]{\textbf{(a)}}\hfill
\makebox[0.49\textwidth]{\textbf{(b)}}
\caption{Conventional plots of the SLAC data.
Panel (a) shows the data with the ${\sim}4\%$ downward
normalization shift applied to the
$1.6~\mathrm{GeV}/c$ spectrometer subset, as in the original
analysis. Panel (b) shows the data without the normalization
adjustment. See text for discussion.}
\label{slac}
\end{figure*}

\section{Re-analysis of 1994 SLAC Data}
Using the Hofstadter-Herman method, graphical representations were constructed using the published cross sections and kinematic SLAC data from Andivahis et al. 1994 \cite{Andivahis:1994}.
The graphs were generated using Python with the NumPy and Matplotlib libraries. As an initial validation step, the reduced cross section graph presented in the original SLAC publication was reproduced. This confirmed the accuracy and reliability of the computational framework. The reproduction was achieved with high fidelity, exhibiting only small deviations, which are addressed in the following subsection.
Following this verification, the code was adapted to express the algebraic relation between $G_E^2(Q^2)$ and $G_M^2(Q^2)$, which was then used to generate Hofstadter-Herman plots. Each band in these plots corresponds to one measurement at a given beam energy and scattering angle, with the band width reflecting the experimental uncertainty in $\sigma_R$. The intersection of multiple bands identifies the consistent pair of form factor values for that $Q^2$. This approach enables direct visualization of the form factors’ behavior, eliminating the need to infer results from linear fits.

\subsection*{Graphical Analysis}
We now apply this framework to the 1994 SLAC dataset. The cross section graph originally published by SLAC was successfully reconstructed, reproducing the overall structure of the 1994 result with minor deviations. These small differences arise from a normalization correction applied to a subset of the data.  According to the SLAC report, reduced cross sections from the 1.6 GeV/$c$ spectrometer were adjusted downward by approximately 4\% (multiplied by $0.958\pm 0.007$ \cite{Andivahis:1994}) to ensure consistency with data collected with the higher energy spectrometers. Panel (a) of Figure~\ref{slac} shows the reconstructed SLAC plot including this normalization adjustment factor.

For comparison, Panel (b) of Figure~\ref{slac} displays the same dataset before normalization. Although the visual differences between the two plots are modest, they illustrate how even small normalization shifts can influence the apparent consistency among measurements. Applying the SLAC-recommended normalization yields a reconstructed graph in close agreement with the published result.

\begin{figure*}[t]
\centering
\includegraphics[width=0.49\textwidth]{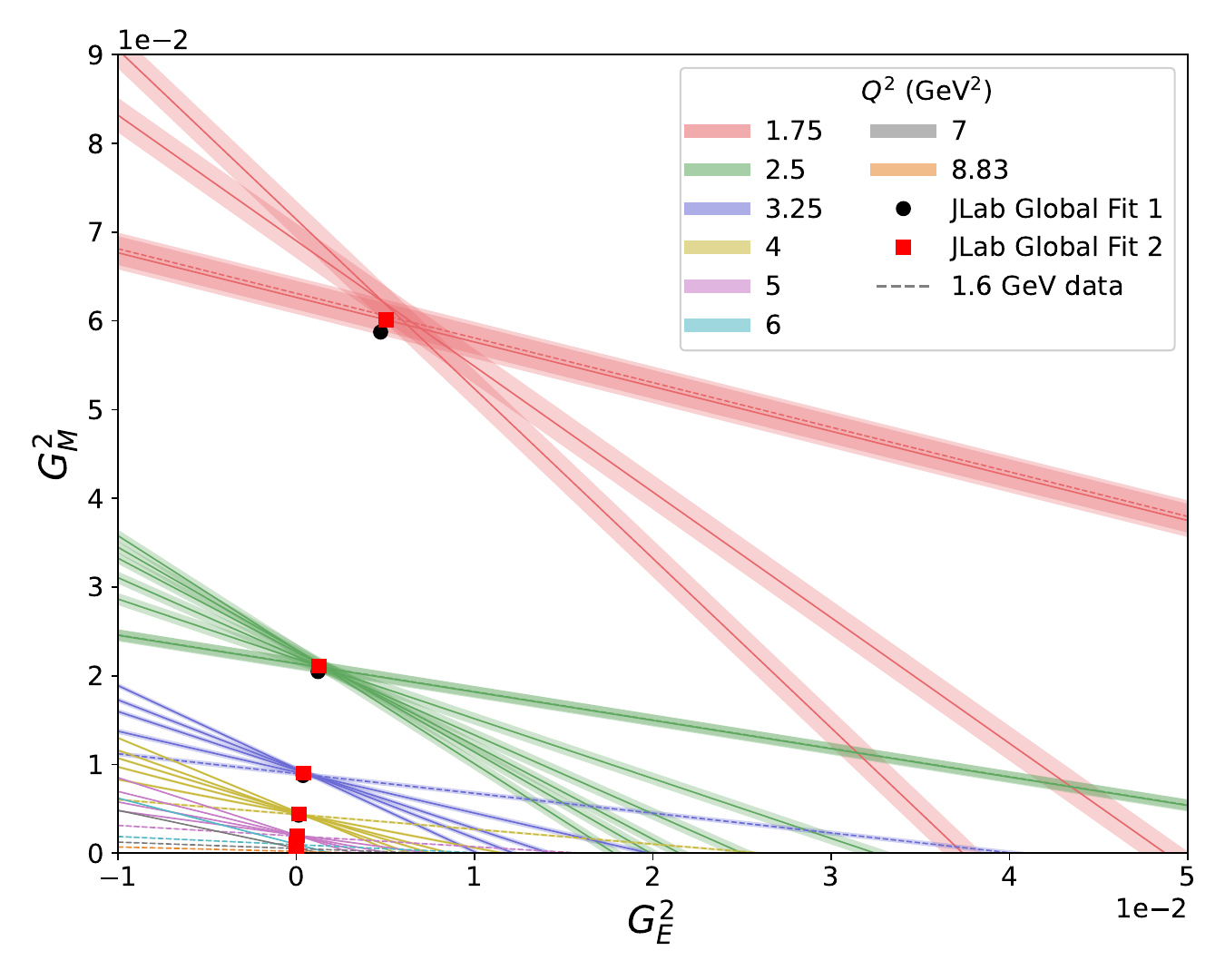}\hfill
\includegraphics[width=0.49\textwidth]{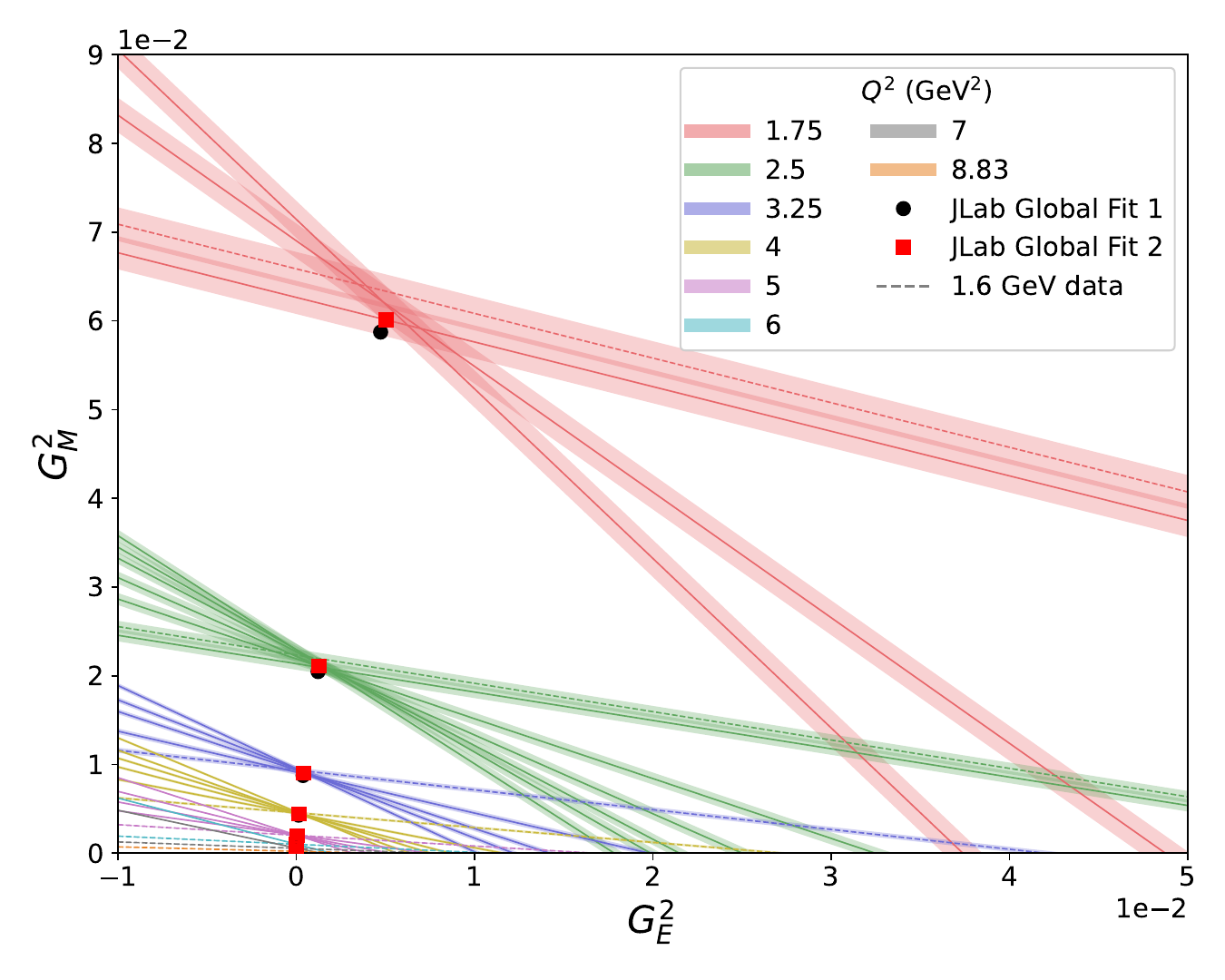}
\makebox[0.49\textwidth]{\textbf{(a)}}\hfill
\makebox[0.49\textwidth]{\textbf{(b)}}
\caption{Hofstadter-Herman plots of the SLAC data.
Panel (a) shows the normalization-adjusted data, and
panel (b) shows the unadjusted data.}
\label{hofstadter}
\end{figure*}

The same dataset was then reanalyzed using the Hofstadter-Herman visualization technique, producing two distinct graphical representations: one incorporating the approximately 4\% correction (Panel (a), Figure \ref{hofstadter}) and one based on the unadjusted data (Panel (b), Figure \ref{hofstadter}). The normalization adjustment shifts the 1.6 $\mathrm{GeV} / \mathrm{c}$ cross sections downward, and this decreases $G_M^2$ and increases $G_E^2$ relative to the unnormalized intersection. The displacement is small, but because $G_E^2$ is itself small, the same absolute shift represents a larger fractional change in the extracted electric contribution. The normalization choice, applied to a single spectrometer's data, is enough to noticeably shift the extracted form factor values away from the global fits, with the electric form factor affected more. A visual comparison of both adjusted and unadjusted data together with the Jefferson Lab global fits (Appendix A of \cite{Puckett:2017}) indicates that the global fit extractions align more closely with the unadjusted data, suggesting that the original uncorrected measurements may better reflect the underlying physical behavior.

Table \ref{shifts} quantifies the visual shifts. At every $Q^2$ where an intersection point is defined, the unadjusted data yields an extracted $\left(G_M^2, G_E^2\right)$ closer to both global fits than the normalization-adjusted data, when distance is measured as the fractional deviation on both form factors relative to the fit value. The fractional separation distance is defined as
\begin{equation}
d_{\mathrm{frac}} = \sqrt{\left(\frac{G_E^2 - G_{E,\mathrm{fit}}^2}{G_{E,\mathrm{fit}}^2}\right)^2 + \left(\frac{G_M^2 - G_{M,\mathrm{fit}}^2}{G_{M,\mathrm{fit}}^2}\right)^2}.
\label{eq:frac_dist}
\end{equation}

\begin{table}[htbp]
\centering
\begin{tabular}{c cc cc c}
\toprule
& \multicolumn{2}{c}{JLab Fit 1} & \multicolumn{2}{c}{JLab Fit 2} & \\
\cmidrule(lr){2-3} \cmidrule(lr){4-5}
$Q^2$ (GeV$^2$) & Unadj. & Adj. & Unadj. & Adj. & Closer \\
\midrule
1.75 & 0.103   & 0.320   & 0.033    & 0.236   & Unadjusted \\
2.50 & 0.055    & 0.404   & 0.070    & 0.351   & Unadjusted \\
3.25 & 0.114   & 1.19  & 0.137   & 1.11  & Unadjusted \\
4.00 & 0.117   & 1.98  & 0.042    & 1.79  & Unadjusted \\
5.00 & 0.172   & 3.50  & 0.272   & 2.93  & Unadjusted \\
6.00 & 0.133   & 6.98  & 0.317   & 5.23  & Unadjusted \\
7.00 & 15.8 & 28.4 & 10.4 & 18.9 & Unadjusted \\
8.83 & --       & --       & --       & --       & N/A \\
\bottomrule
\end{tabular}
\caption{Fractional separation distance between the Hofstadter-Herman intersection point and the JLab global fits for the unadjusted and normalization-adjusted (of the 1.6 GeV/$c$ spectrometer subset) data. The fractional distance formula is given in Eq.~\ref{eq:frac_dist}. At every $Q^2$ where an intersection exists, the unadjusted point is closer to both JLab global fit values.}
\label{shifts}
\end{table}

The tensions observed between the 1994 data and the 2017 Jefferson Lab global fit extractions are difficult to see in conventional Rosenbluth separation plots. By contrast, the Hofstadter-Herman visualization shows the regions of agreement as overlapping bands of allowed form factor values at fixed $Q^{2}$ across different beam energies.  In this representation, the consistent form factor values correspond to the region where the bands intersect, and the JLab Global Fits can be seen to pass through or deviate from this region. As stated in the 8-panel figure captions, the intersection markers sit at the bandwidth-weighted least-squares common point of the central constraint lines. The results from this visualization method underscore the potential limitations of relying solely on global fits without reference to the underlying experimental distributions.

\begin{figure*}[!htbp]
    \includegraphics[width =0.9\textwidth]{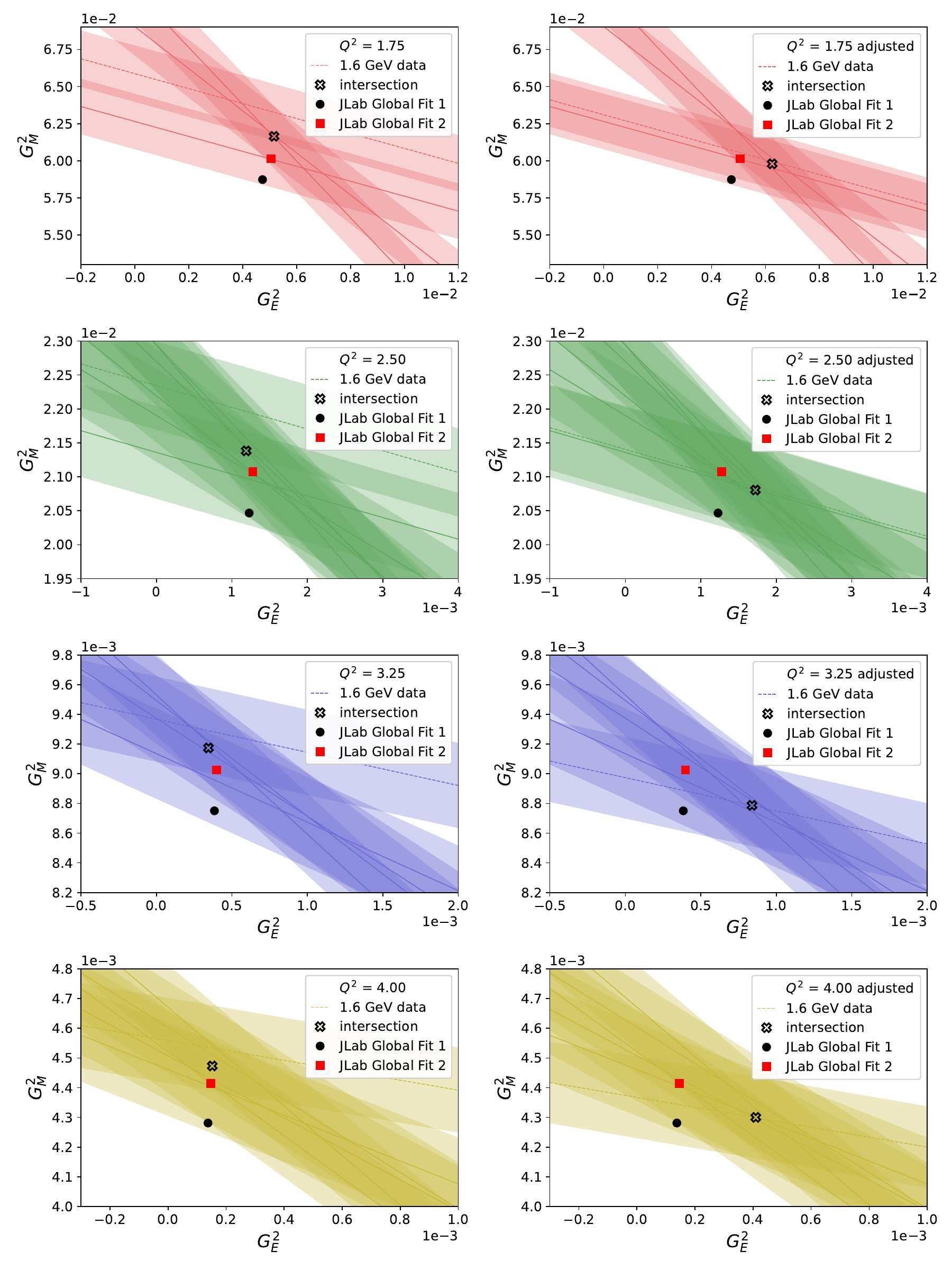}
   \caption{An 8-panel Hofstadter-Herman display of SLAC data for the lower range of $Q^2$ values (in units of $\rm GeV^2$). The left side is the unmodified data, and the right side has the $\sim$4\% normalization adjustment applied to the 1.6 GeV/$c$ spectrometer data subset, as in the original 1994 SLAC analysis. The intersection markers near the center of the overlap region are points that are consistent with all $(G_M^2, G_E^2)$ values and are computed by finding the bandwidth-weighted least-squares distance to the central constraint lines.}
\end{figure*}

\begin{figure*}[!htbp]
    \includegraphics[width =0.9\textwidth]{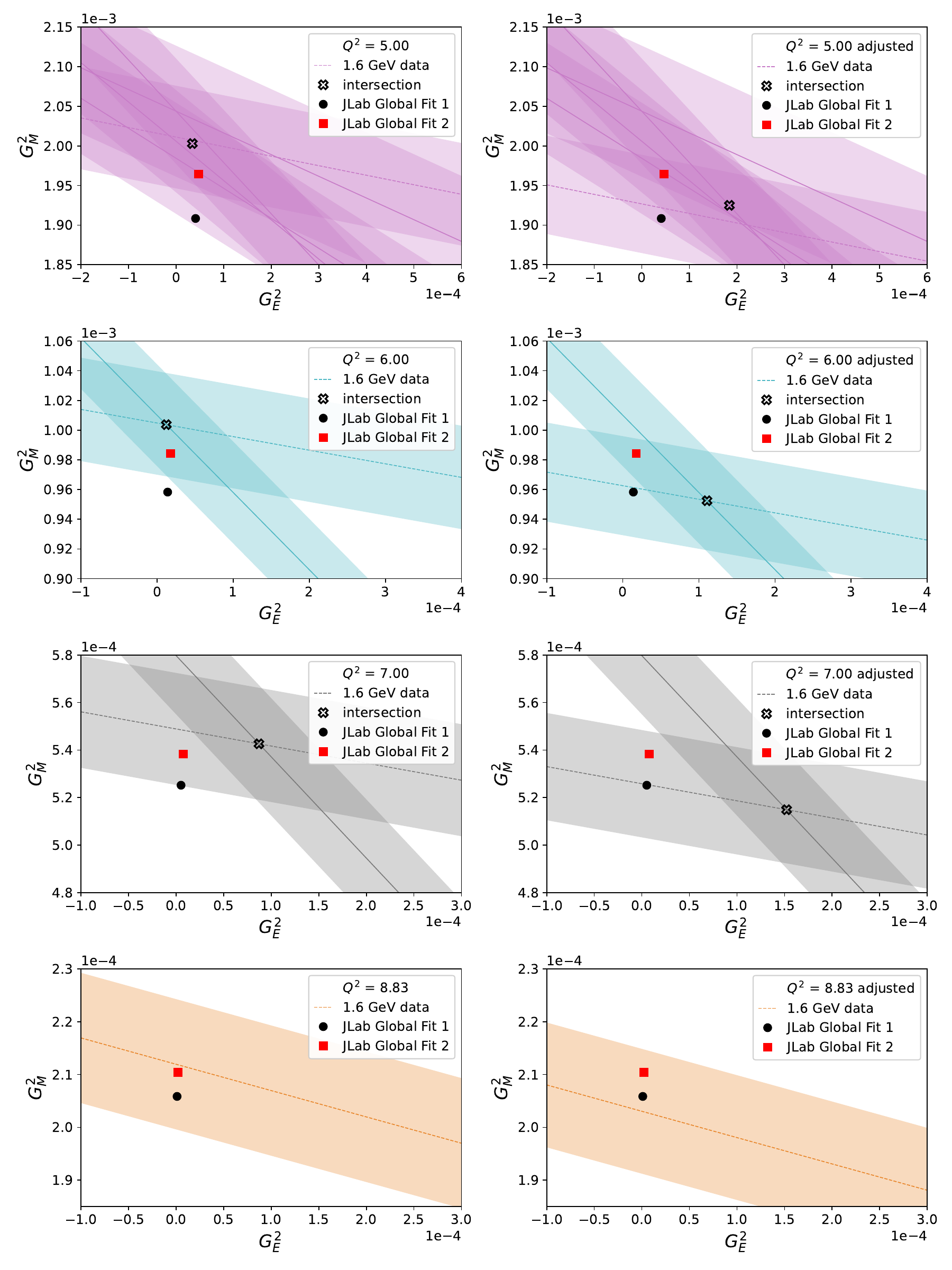}
   \caption{An 8-panel Hofstadter-Herman display of SLAC data for the higher range of $Q^2$ values (in units of $\rm GeV^2$). The left side is the unmodified data, and the right side has the $\sim$4\% normalization adjustment applied to the 1.6 GeV/$c$ spectrometer data subset, as in the original 1994 SLAC analysis. The final row in orange shows only one band because SLAC had only one measurement at $Q^2=8.83$ GeV$^2$. The intersection markers near the center of the overlap region are points that are consistent with all $(G_M^2, G_E^2)$ values and are computed by finding the bandwidth-weighted least-squares distance to the central constraint lines.}
    \label{Q2-zoom}
\end{figure*}

\section{Future Applications and Conclusions}
We have revisited classic SLAC elastic $ep$ scattering data using both conventional Rosenbluth separation plots and the Hofstadter-Herman visualization method.  The latter presents the form factors directly in $(G_M^2,G_E^2)$ space rather than through the usual two-step procedure of fitting a line to the data and then extracting $G_E^2$ and $G_M^2$ from the slope and intercept. This approach reveals normalization inconsistencies that are difficult to see in traditional plots.  In the SLAC dataset, a $\sim$4\% downward normalization correction applied to the 1.6 GeV/$c$ spectrometer data in the original SLAC analysis shifted measurements away from the region favored by modern global fits, an effect seen in the Hofstadter-Herman plots but obscured in conventional plots. The shift suggests that the uncorrected data may more accurately reflect the underlying behavior of the form factors.  

By mapping the allowed $(G_E^2, G_M^2)$ regions directly, Hofstadter–Herman visualization can make certain inconsistencies between data sets apparent, particularly those arising from normalization or calibration shifts that may remain hidden in conventional plots. Such geometric cross-checks can complement standard analysis techniques at current and future facilities, including the Electron–Ion Collider, by helping identify normalization offsets or other biases before they propagate into global fits. More broadly, this work underscores the continuing value of geometry-based representations of form factor data for assessing the internal consistency of high-precision measurements.

\section{Acknowledgements}
T.W and F.B acknowledge support from the National Science Foundation, Award No.Benmokhtar-2310067 and No.Benmokhtar-2610003. J.R.W. acknowledges support from the Istituto Nazionale di Fisica Nucleare (INFN) Turin, Grant No. 25864. This research was funded in part by the Department of Energy grant number DE-AC05-06OR23177.

\appendix 
\section{Error analysis}
  In this section, details of the error analysis are presented. We explain the normalization of the reduced cross section and the associated errors, then the point-to-point systematic errors. Finally, we give a breakdown of the overall normalization uncertainty.

    \subsection{ \textbf{Normalization Adjustment of Reduced Cross Sections}}
    \begin{enumerate}
        \item The reduced cross sections from the 1.6 GeV/$c$ spectrometer were normalized to the measurements of the 8 GeV/$c$ spectrometer \cite{Andivahis:1994}. Normalization factors were obtained by fitting the 8 GeV/$c$ reduced cross sections as a function of the virtual photon polarization $\epsilon$ at the five lowest $Q^{2}$ values.
        \item Only $Q^{2}$ points with at least two 8 GeV data points were included to allow a linear fit. For each $Q^{2}$, a normalization factor was extracted to scale the 1.6 GeV/$c$ reduced cross section to the fitted line.
        \item The five normalization factors were consistent with being independent of $Q^{2}$. The factor $0.958 \pm 0.007$, obtained at the lowest $Q^{2}$, was applied to all $Q^{2}$ values.
        \item This $\sim$4\% deviation from unity was attributed to uncertainty in the 1.6 GeV/$c$ acceptance function.
        \item An additional systematic uncertainty of $\pm 0.7\%$ due to this $\sim$4\% normalization adjustment was assigned to the 1.6 GeV/$c$ reduced cross sections.
    \end{enumerate}

    \subsection{\textbf{Point-to-Point Systematic Errors}}  
Systematic uncertainties were obtained by adding in quadrature the point-to-point uncertainties from:
\vspace{0.2cm}
        \begin{enumerate}
            \item Incident charge: 0.2\%
            \item Target density: 0.2\%
            \item Detector efficiency: 0.2\% (8 GeV/$c$), 0.3\% (1.6 GeV/$c$)
            \item Electronic and computer dead time: 0.2\%
            \item Beam energy: 0.06\%
            \item Scattering angle: $0.006^\circ$ (8 GeV/$c$), $0.05^\circ$ (1.6 GeV/$c$)
            \item Background pion subtraction: negligible
            \item Aluminum end-cap subtraction: 0.2\%
            \item Radiative corrections: 0.5\%
            \item Spectrometer acceptance: 0.5\% (8 GeV/$c$), 0.75\% (1.6 GeV/$c$)
            \item Normalization uncertainty for 1.6 GeV/$c$: 0.7\%
        \end{enumerate}
        \vspace{0.2cm}
        Quadrature sums resulted in 1.06\%  and 1.32\% point-to-point systematic uncertainty for the 8 GeV/$c$ and 1.6 GeV/$c$ setups, respectively. 
         
        \vspace{0.2cm}
         Finally, the total point-to-point uncertainty is obtained by adding statistical uncertainty ($\approx$$1\%$) in quadrature with systematic uncertainty.

    \subsection{\textbf{Overall Normalization Uncertainty}}   
    A separate overall normalization uncertainty of $\pm$1.77\% was applied to all reduced cross sections \cite{Andivahis:1994}. This uncertainty is the quadrature sum of the incident charge: 0.5\%, target density: 0.9\%, beam energy: 0.05\%, radiative corrections: 1.0\%, and 8 GeV acceptance function: 1.0\% (values rounded). 
    
    This uncertainty is not included in the tables but is included in the uncertainty bands in the plots. The band half-widths in the plots are the tabulated total error plus a 1.77\% normalization uncertainty added linearly to each band independently. This is a deliberately wide envelope for the intersection test; it is not a confidence region, and it does not treat the normalization as the single experiment-wide scale factor that it is.

\clearpage
\newpage
\begin{table*}[htp]
\begin{center}
\begin{tabular}{c|c|c|c|c|c|c|c}
$Q^2$ & E & $\theta$ & E` & $\epsilon$ & $\frac{d\sigma}{d\Omega}$ & $\pm$Stat. error & $\pm$Total error \\\hline
1.75 & 1.511 & 90.066 & 0.578 & 0.250 & 1.440 x 10\textsuperscript{-1} & 1.116 x 10\textsuperscript{-3} & 1.750 x 10\textsuperscript{-3} \\
1.75 & 2.407 & 41.110 & 1.474 & 0.704 & 1.029 x 10\textsuperscript{0} & 4.715 x 10\textsuperscript{-3} & 1.090 x 10\textsuperscript{-2} \\
1.75 & 5.507 & 15.145 & 4.574 & 0.950 & 1.155 x 10\textsuperscript{+1} & 6.713 x 10\textsuperscript{-2} & 1.336 x 10\textsuperscript{-1} \\
2.50 & 1.968 & 89.947 & 0.636 & 0.227 & 3.389 x 10\textsuperscript{-2} & 3.616 x 10\textsuperscript{-4} & 4.832 x 10\textsuperscript{-4} \\
2.50 & 2.407 & 58.882 & 1.075 & 0.479 & 9.857 x 10\textsuperscript{-2} & 9.199 x 10\textsuperscript{-4} & 1.317 x 10\textsuperscript{-3} \\
2.50 & 2.837 & 44.993 & 1.505 & 0.630 & 1.990 x 10\textsuperscript{-1} & 1.811 x 10\textsuperscript{-3} & 2.638 x 10\textsuperscript{-3} \\
2.50 & 3.400 & 34.694 & 2.068 & 0.750 & 3.951 x 10\textsuperscript{-1} & 1.849 x 10\textsuperscript{-3} & 4.266 x 10\textsuperscript{-3} \\
2.50 & 3.956 & 28.409 & 2.624 & 0.820 & 6.616 x 10\textsuperscript{-1} & 4.025 x 10\textsuperscript{-3} & 7.637 x 10\textsuperscript{-3} \\
2.50 & 5.507 & 18.981 & 4.175 & 0.913 & 1.779 x 10\textsuperscript{0} & 1.147 x 10\textsuperscript{-2} & 2.120 x 10\textsuperscript{-2} \\
3.25 & 2.837 & 61.205 & 1.105 & 0.426 & 2.848 x 10\textsuperscript{-2} & 3.499 x 10\textsuperscript{-4} & 4.444 x 10\textsuperscript{-4} \\
3.25 & 3.400 & 44.482 & 1.668 & 0.609 & 6.784 x 10\textsuperscript{-2} & 5.949 x 10\textsuperscript{-4} & 8.885 x 10\textsuperscript{-4} \\
3.25 & 3.956 & 35.382 & 2.224 & 0.719 & 1.256 x 10\textsuperscript{-1} & 1.075 x 10\textsuperscript{-3} & 1.636 x 10\textsuperscript{-3} \\
3.25 & 5.507 & 22.804 & 3.775 & 0.865 & 3.898 x 10\textsuperscript{-1} & 1.888 x 10\textsuperscript{-3} & 4.343 x 10\textsuperscript{-3} \\
4.00 & 3.400 & 57.572 & 1.268 & 0.437 & 1.297 x 10\textsuperscript{-2} & 1.858 x 10\textsuperscript{-4} & 2.243 x 10\textsuperscript{-4} \\
4.00 & 3.956 & 43.707 & 1.824 & 0.593 & 2.770 x 10\textsuperscript{-2} & 3.474 x 10\textsuperscript{-4} & 4.407 x 10\textsuperscript{-4} \\
4.00 & 4.507 & 35.592 & 2.375 & 0.694 & 4.929 x 10\textsuperscript{-2} & 6.162 x 10\textsuperscript{-4} & 7.853 x 10\textsuperscript{-4} \\
4.00 & 5.507 & 26.823 & 3.375 & 0.805 & 1.023 x 10\textsuperscript{-1} & 9.097 x 10\textsuperscript{-4} & 1.370 x 10\textsuperscript{-3} \\
4.00 & 9.800 & 13.248 & 7.668 & 0.946 & 6.180 x 10\textsuperscript{-1} & 4.679 x 10\textsuperscript{-3} & 8.073 x 10\textsuperscript{-3} \\
5.00 & 3.956 & 59.291 & 1.291 & 0.389 & 4.205 x 10\textsuperscript{-3} & 8.647 x 10\textsuperscript{-5} & 9.565 x 10\textsuperscript{-5} \\
5.00 & 4.507 & 45.658 & 1.842 & 0.538 & 8.462 x 10\textsuperscript{-3} & 1.239 x 10\textsuperscript{-4} & 1.492 x 10\textsuperscript{-4} \\
5.00 & 5.507 & 32.829 & 2.842 & 0.704 & 2.128 x 10\textsuperscript{-2} & 2.228 x 10\textsuperscript{-4} & 3.079 x 10\textsuperscript{-4} \\
5.00 & 9.800 & 15.367 & 7.135 & 0.919 & 1.576 x 10\textsuperscript{-1} & 1.643 x 10\textsuperscript{-3} & 2.338 x 10\textsuperscript{-3} \\
6.00 & 9.800 & 17.515 & 6.603 & 0.886 & 4.749 x 10\textsuperscript{-2} & 5.879 x 10\textsuperscript{-4} & 7.705 x 10\textsuperscript{-4} \\
7.00 & 9.800 & 19.753 & 6.070 & 0.847 & 1.707 x 10\textsuperscript{-2} & 3.860 x 10\textsuperscript{-4} & 4.249 x 10\textsuperscript{-4} \\
\end{tabular}
\caption{\label{tab:widgets8}Data provided by the 8 GeV spectrometer at SLAC. We note that row 5 has been corrected from the SLAC 1994 data table IV, $\epsilon=0.479$ not $0.47$. We also note that row 8 has the correct $3.956$ value for E, not the original $3.95$ value from the SLAC 1994 data table IV. Elastic scattering kinematic equations confirm both corrections.}
\end{center}
\end{table*}

\begin{table*}[htp]
\centering
\begin{tabular}{c|c|c|c|c|c|c|c}
$Q^2$ & E & $\theta$ & E` & $\epsilon$ & $\frac{d\sigma}{d\Omega}$ & $\pm$Stat. error & $\pm$Total error \\\hline
1.75 & 1.511 & 90.066 & 0.578 & 0.250 & 1.514 x 10\textsuperscript{-1} & 3.132 x 10\textsuperscript{-4} & 1.690 x 10\textsuperscript{-3} \\
2.50 & 1.968 & 89.947 & 0.636 & 0.227 & 3.545 x 10\textsuperscript{-2} & 1.008 x 10\textsuperscript{-4} & 4.044 x 10\textsuperscript{-4} \\
3.25 & 2.407 & 90.004 & 0.675 & 0.206 & 1.095 x 10\textsuperscript{-2} & 7.314 x 10\textsuperscript{-5} & 1.418 x 10\textsuperscript{-4} \\
4.00 & 2.837 & 89.966 & 0.705 & 0.190 & 4.092 x 10\textsuperscript{-3} & 3.323 x 10\textsuperscript{-5} & 5.636 x 10\textsuperscript{-5} \\
5.00 & 3.400 & 89.985 & 0.735 & 0.171 & 1.339 x 10\textsuperscript{-3} & 1.242 x 10\textsuperscript{-5} & 1.942 x 10\textsuperscript{-5} \\
6.00 & 3.956 & 89.981 & 0.759 & 0.156 & 5.164 x 10\textsuperscript{-4} & 6.577 x 10\textsuperscript{-6} & 8.747 x 10\textsuperscript{-6} \\
7.00 & 4.507 & 89.991 & 0.777 & 0.143 & 2.248 x 10\textsuperscript{-4} & 5.088 x 10\textsuperscript{-6} & 5.675 x 10\textsuperscript{-6} \\
8.83 & 5.507 & 90.016 & 0.801 & 0.125 & 6.022 x 10\textsuperscript{-5} & 2.344 x 10\textsuperscript{-6} & 2.439 x 10\textsuperscript{-6} \\
\end{tabular}
\caption{\label{tab:widgets1.6}Data provided by the 1.6 GeV spectrometer at SLAC. We note that row 8 has the correct $0.801$ value for E', not the original $0.784$ value from the SLAC 1994 data table IV. Elastic scattering kinematic equations confirm the correction.}
\end{table*}

\newpage

\begin{table*}[htp]
\begin{center}
\begin{tabular}
{c|c|c}
$Q^2$ & $G_M^2$ & $G_E^2$ \\\hline 1.75 & 5.8733 x 10\textsuperscript{-2} & 4.7353 x 10\textsuperscript{-3} \\ 2.50 & 2.0467 x 10\textsuperscript{-2} & 1.2284 x 10\textsuperscript{-3} \\ 3.25 & 8.7506 x 10\textsuperscript{-3} & 3.8403 x 10\textsuperscript{-4} \\ 4.00 & 4.2811 x 10\textsuperscript{-3} & 1.3751 x 10\textsuperscript{-4} \\ 5.00 & 1.9084 x 10\textsuperscript{-3} & 4.0886 x 10\textsuperscript{-5} \\ 6.00 & 9.5830 x 10\textsuperscript{-4} & 1.3866 x 10\textsuperscript{-5} \\ 7.00 & 5.2518 x 10\textsuperscript{-4} & 5.1696 x 10\textsuperscript{-6} \\ 8.83 & 2.0585 x 10\textsuperscript{-4} &  9.9541 x 10\textsuperscript{-7} \\
\end{tabular}
\caption{\label{tab:FitList}The plotted fit data based on JLab Global Fit \small{$\#$}1 \cite{Puckett:2017}, with $\mu_p=2.793$, $M_p=0.938$ GeV used in the form factor parameterization equation. \cite{Puckett:2017,Gramolin:2016hjt}}
\end{center}
\end{table*}

\begin{table*}[htp]
\begin{center}
\begin{tabular}
{c|c|c}
$Q^2$ & $G_M^2$ & $G_E^2$ \\\hline
1.75 & 6.0135 x 10\textsuperscript{-2}  & 5.0558 x 10\textsuperscript{-3} \\ 2.50 & 2.1076 x 10\textsuperscript{-2} & 1.2761 x 10\textsuperscript{-3} \\ 3.25 & 9.0264 x 10\textsuperscript{-3} & 3.9838 x 10\textsuperscript{-4} \\ 4.00 & 4.4147 x 10\textsuperscript{-3} & 1.4652 x 10\textsuperscript{-4} \\ 5.00 & 1.9645 x 10\textsuperscript{-3} & 4.6895 x 10\textsuperscript{-5} \\ 6.00 & 9.8425 x 10\textsuperscript{-4} & 1.7757 x 10\textsuperscript{-5} \\ 7.00 & 5.3828 x 10\textsuperscript{-4} & 7.6354 x 10\textsuperscript{-6} \\ 8.83 & 2.1043 x 10\textsuperscript{-4} & 2.0741 x 10\textsuperscript{-6} \\
\end{tabular}
\caption{\label{tab:FitList2}The plotted fit data based on JLab Global Fit 2 \cite{Puckett:2017}, with $\mu_p=2.793$, $M_p=0.938$ GeV used in the form factor parameterization equation. \cite{Puckett:2017,Gramolin:2016hjt}}
\end{center}
\end{table*}

\bibliographystyle{ieeetr}

\bibliography{references}

@article{Andivahis:1994,
    author = "Andivahis, L. and others",
    title = "Measurements of the electric and magnetic form factors of the proton from $Q^2=1.75 to 8.83 (GeV/c)^2$",
    journal = "Phys. Rev. D",
    volume = "50",
    year = "1994",
    pages = "5491"
}

@article{Gramolin:2016hjt,
    author = "Gramolin, A. V. and Nikolenko, D. M.",
    title = "{Reanalysis of Rosenbluth measurements of the proton form factors}",
    eprint = "1603.06920",
    archivePrefix = "arXiv",
    primaryClass = "nucl-ex",
    doi = "10.1103/PhysRevC.93.055201",
    journal = "Phys. Rev. C",
    volume = "93",
    number = "5",
    pages = "055201",
    year = "2016"
}

@book{Hofstadter:1960,
    author = "Herman, R. and Hofstadter, R.",
    title = "High-Energy Electron Scattering Tables",
    publisher = "Stanford University Press",
    year = "1960",
    pages = "31"
}

@article{Arrington:2011kb,
    author = "Arrington, John and de Jager, Kees and Perdrisat, Charles F.",
    title = "{Nucleon Form Factors: A Jefferson Lab Perspective}",
    eprint = "1102.2463",
    archivePrefix = "arXiv",
    primaryClass = "nucl-ex",
    reportNumber = "JLAB-PHY-11-1315",
    doi = "10.1088/1742-6596/299/1/012002",
    journal = "J. Phys. Conf. Ser.",
    volume = "299",
    pages = "012002",
    year = "2011"
}

@article{Puckett:2017,
    author = "Puckett, A. J. R. and others",
    title = "{Polarization Transfer Observables in Elastic Electron Proton Scattering at $Q^2 = $2.5, 5.2, 6.8, and 8.5 GeV$^2$}",
    eprint = "1707.08587",
    archivePrefix = "arXiv",
    primaryClass = "nucl-ex",
    reportNumber = "JLAB-PHY-17-2533",
    doi = "10.1103/PhysRevC.96.055203",
    journal = "Phys. Rev. C",
    volume = "96",
    number = "5",
    pages = "055203",
    year = "2017",
    note = "[Erratum: Phys.Rev.C 98, 019907 (2018)]"
}

@article{Hand:1963zz,
    author = "Hand, L. N. and Miller, D. G. and Wilson, Richard",
    title = "{Electric and Magnetic Form Factors of the Nucleon}",
    doi = "10.1103/RevModPhys.35.335",
    journal = "Rev. Mod. Phys.",
    volume = "35",
    pages = "335",
    year = "1963"
}

@article{Kubon:2001rj,
    author = "Kubon, G. and others",
    title = "{Precise neutron magnetic form-factors}",
    eprint = "nucl-ex/0107016",
    archivePrefix = "arXiv",
    doi = "10.1016/S0370-2693(01)01386-7",
    journal = "Phys. Lett. B",
    volume = "524",
    pages = "26--32",
    year = "2002"
}

@article{Ye:2017gyb,
    author = "Ye, Zhihong and Arrington, John and Hill, Richard J. and Lee, Gabriel",
    title = "{Proton and Neutron Electromagnetic Form Factors and Uncertainties}",
    eprint = "1707.09063",
    archivePrefix = "arXiv",
    primaryClass = "nucl-ex",
    reportNumber = "FERMILAB-PUB-17-281-T",
    doi = "10.1016/j.physletb.2017.11.023",
    journal = "Phys. Lett. B",
    volume = "777",
    pages = "8--15",
    year = "2018"
}

@article{Alberico:2008sz,
    author = "Alberico, W. M. and Bilenky, S. M. and Giunti, C. and Graczyk, K. M.",
    title = "{Electromagnetic form factors of the nucleon: New Fit and analysis of uncertainties}",
    eprint = "0812.3539",
    archivePrefix = "arXiv",
    primaryClass = "hep-ph",
    doi = "10.1103/PhysRevC.79.065204",
    journal = "Phys. Rev. C",
    volume = "79",
    pages = "065204",
    year = "2009"
}

@article{Lin:2021xrc,
    author = "Lin, Yong-Hui and Hammer, Hans-Werner and Mei{\ss}ner, Ulf-G.",
    title = "{New Insights into the Nucleon{\textquoteright}s Electromagnetic Structure}",
    eprint = "2109.12961",
    archivePrefix = "arXiv",
    primaryClass = "hep-ph",
    doi = "10.1103/PhysRevLett.128.052002",
    journal = "Phys. Rev. Lett.",
    volume = "128",
    number = "5",
    pages = "052002",
    year = "2022"
}

@article{A1:2013fsc,
    author = "Bernauer, J. C. and others",
    collaboration = "A1",
    title = "{Electric and magnetic form factors of the proton}",
    eprint = "1307.6227",
    archivePrefix = "arXiv",
    primaryClass = "nucl-ex",
    doi = "10.1103/PhysRevC.90.015206",
    journal = "Phys. Rev. C",
    volume = "90",
    number = "1",
    pages = "015206",
    year = "2014"
}

@article{Hofstadter:1955ae,
    author = "Hofstadter, R. and McAllister, R. W.",
    title = "{Electron Scattering From the Proton}",
    doi = "10.1103/PhysRev.98.217",
    journal = "Phys. Rev.",
    volume = "98",
    pages = "217--218",
    year = "1955"
}

@article{Rosenbluth:1950yq,
    author = "Rosenbluth, M. N.",
    title = "{High Energy Elastic Scattering of Electrons on Protons}",
    doi = "10.1103/PhysRev.79.615",
    journal = "Phys. Rev.",
    volume = "79",
    pages = "615--619",
    year = "1950"
}

@article{Sachs:1962zzc,
    author = "Sachs, R. G.",
    title = "{High-Energy Behavior of Nucleon Electromagnetic Form Factors}",
    doi = "10.1103/PhysRev.126.2256",
    journal = "Phys. Rev.",
    volume = "126",
    pages = "2256--2260",
    year = "1962"
}

@article{Perdrisat:2006hj,
    author = "Perdrisat, C. F. and Punjabi, V. and Vanderhaeghen, M.",
    title = "{Nucleon Electromagnetic Form Factors}",
    eprint = "hep-ph/0612014",
    archivePrefix = "arXiv",
    reportNumber = "WM-06-115, JLAB-THY-06-595",
    doi = "10.1016/j.ppnp.2007.05.001",
    journal = "Prog. Part. Nucl. Phys.",
    volume = "59",
    pages = "694--764",
    year = "2007"
}

@article{Accardi:2012qut,
    author = "Accardi, A. and others",
    editor = "Deshpande, A. and Meziani, Z. E. and Qiu, J. W.",
    title = "{Electron Ion Collider: The Next QCD Frontier}: {Understanding the glue that binds us all}",
    eprint = "1212.1701",
    archivePrefix = "arXiv",
    primaryClass = "nucl-ex",
    reportNumber = "BNL-98815-2012-JA, JLAB-PHY-12-1652",
    doi = "10.1140/epja/i2016-16268-9",
    journal = "Eur. Phys. J. A",
    volume = "52",
    number = "9",
    pages = "268",
    year = "2016"
}

@article{Achenbach:2023pba,
    author = "Achenbach, P. and others",
    title = "{The present and future of QCD}",
    eprint = "2303.02579",
    archivePrefix = "arXiv",
    primaryClass = "hep-ph",
    reportNumber = "JLAB-PHY-23-3808",
    doi = "10.1016/j.nuclphysa.2024.122874",
    journal = "Nucl. Phys. A",
    volume = "1047",
    pages = "122874",
    year = "2024"
}

@article{Rose:1948zz,
    author = "Rose, M. E.",
    title = "{The Charge Distribution in Nuclei and the Scattering of High Energy Electrons}",
    doi = "10.1103/PhysRev.73.279",
    journal = "Phys. Rev.",
    volume = "73",
    pages = "279--284",
    year = "1948"
}

@article{Higinbotham:2015rja,
    author = "Higinbotham, Douglas W. and Kabir, Al Amin and Lin, Vincent and Meekins, David and Norum, Blaine and Sawatzky, Brad",
    title = "{Proton radius from electron scattering data}",
    eprint = "1510.01293",
    archivePrefix = "arXiv",
    primaryClass = "nucl-ex",
    reportNumber = "JLAB-PHY-16-2",
    doi = "10.1103/PhysRevC.93.055207",
    journal = "Phys. Rev. C",
    volume = "93",
    number = "5",
    pages = "055207",
    year = "2016"
}

@article{Karr:2020wgh,
    author = "Karr, Jean-Philippe and Marchand, Dominique and Voutier, Eric",
    title = "{The proton size}",
    doi = "10.1038/s42254-020-0229-x",
    journal = "Nature Rev. Phys.",
    volume = "2",
    number = "11",
    pages = "601--614",
    year = "2020"
}

\end{document}